\documentclass[prd,twocolumn]{revtex4}
\usepackage{graphicx, epsfig}
\usepackage{color}
\usepackage{mathrsfs}

\newcommand{\be}{\begin{equation}}
\newcommand{\ee}{\end{equation}}
\newcommand{\bea}{\begin{eqnarray}}
\newcommand{\eea}{\end{eqnarray}}

\newcommand{\gapp}{\mathrel{\raise.3ex\hbox{$>$}\mkern-14mu
              \lower0.6ex\hbox{$\sim$}}}
\newcommand{\lapp}{\mathrel{\raise.3ex\hbox{$<$}\mkern-14mu
              \lower0.6ex\hbox{$\sim$}}}
\newcommand{\scri}{\mathscr{I}}

\begin{document}
\title{Gravitational Waves, Gamma Ray Bursts, and Black Stars\footnote{Essay written for the Gravity Research Foundation 2016 Awards for Essays on Gravitation.}}
\author{Tanmay Vachaspati}
\affiliation{
Physics Department, Arizona State University, Tempe, AZ 85287, USA.
\\
E-mail: tvachasp@asu.edu
}

\date{\today}

\begin{abstract}
\noindent

Stars that are collapsing toward forming a black hole but appear frozen near their 
Schwarzschild horizon are termed ``black stars''. The collision of two black stars
leads to gravitational radiation during the merging phase followed by a delayed 
gamma ray burst during coalescence.
The recent observation of gravitational waves by LIGO, followed by a possible
gamma ray counterpart by Fermi, suggests that the source may have been a merger of
two black stars with profound implications for quantum gravity and the nature
of black holes.
\end{abstract}

\maketitle


From an asymptotic observer's viewpoint, a collapsing body is forever 
suspended just above its Schwarzschild radius. The strong gravitational 
redshift near the surface of the collapsing body causes the body
to appear black. Such objects are known as ``frozen stars'' or
``black stars''. Black {\it holes} are the infinite time limit of black stars and
traditionally black stars are viewed as indistinguishable from
black holes ({\it e.g.} Chapter 33, \cite{Misner:1974qy}). However,
there are good reasons to maintain the distinction between black
stars and black holes. First, 
{\it quantum} analyses of gravitational collapse show that a black star
evaporates in a finite time \cite{Vachaspati:2006ki} and so it is
impossible to take the infinite time limit. Second,
theories of quantum gravity often predict that black holes have
structure such as a string theory fuzzball \cite{Mathur:2005zp} or a 
firewall \cite{Almheiri:2012rt}.
Third, observations of the collision of two black objects can tell us if the
objects are black holes or black stars and hence the distinction between 
these objects is experimentally meaningful \cite{Vachaspati:2007fc}.
The resolution of these issues has taken on a new urgency after
the recent exciting observations by LIGO \cite{Abbott:2016blz} and 
Fermi \cite{Connaughton:2016umz} in which gravitational
wave emission from two coalescing black hole-like objects appears to have 
been followed by a gamma ray burst.

%

\begin{figure}
  \includegraphics[height=0.35\textwidth,angle=0]{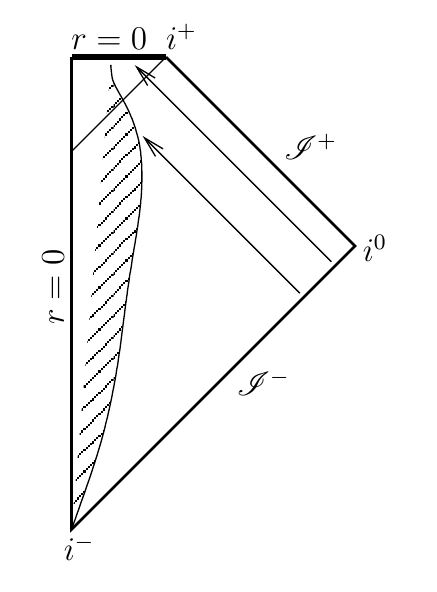}
\caption{Spacetime of a classical collapsing object (shaded region)
that appears as a black star to an observer located near $i^+$. 
Two null rays from $\scri^-$ are shown, one that collides with 
the black star and the other that does not.
}
\label{bh_spacetime}
\end{figure}

The essential idea behind equating black stars to black holes 
is that a collapsing star very quickly fades from an observer's view,
and there is no way to send in probes ({\it e.g.} light rays) at late times so as 
to see the surface of the black star. This idea is captured in the spacetime
diagram shown in Fig.~\ref{bh_spacetime} where certain rays can
hit the surface of the star but later rays arrive at the surface
of the star after it has crossed into its own event horizon. So the
surface of the black star can only be probed by rays that
arrive sufficiently early. This is the usual interpretation, also described in
\cite{Misner:1974qy}, and if the object is probed at sufficiently late times,
there is no way to send in probes to distinguish between a black star and 
a black hole.


\begin{figure}
  \includegraphics[height=0.5\textwidth,angle=0]{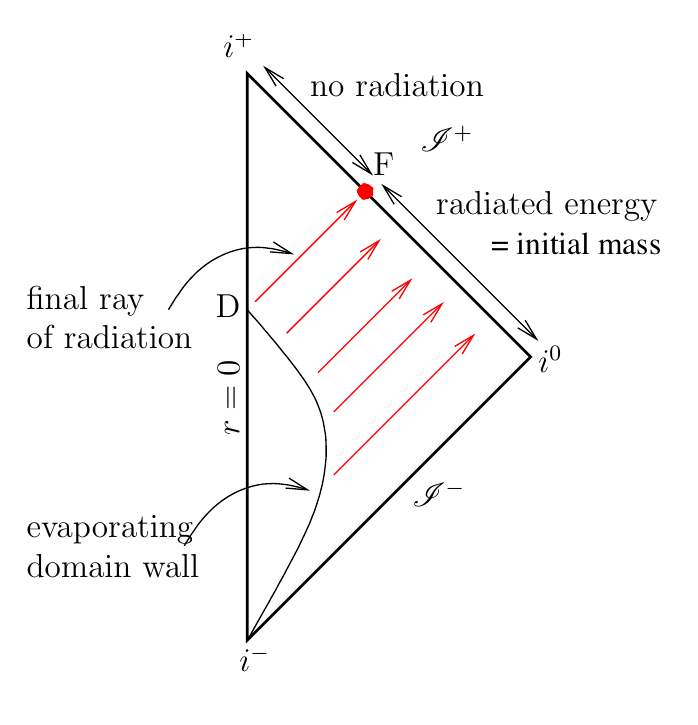}
\caption{Spacetime of a quantum collapsing object 
(labeled a ``domain wall'' in the figure) \cite{Vachaspati:2006ki}.
Now there is no event horizon and singularity, and all rays that
hit the collapsing object reach $\scri^+$, even if they are delayed
due to their interactions with the gravitational field. 
}
\label{qspacetime}
\end{figure}

The picture changes in a {\it quantum} analysis, since then a collapsing 
body produces a time-dependent metric that leads to quantum radiation 
and causes the body to slowly 
evaporate even as it collapses \cite{Vachaspati:2006ki}. This
radiation, very similar to Hawking radiation from a black hole  \cite{Hawking:1974sw},
does not require the event horizon of a black hole and holds for any collapsing body, 
even when it is outside its own Schwarzschild radius. 
Gravitational collapse leads to a black star that 
is continually collapsing and concurrently evaporating into quantum radiation.
Then the collapsing object spacetime is shown in Fig.~\ref{qspacetime}.
Now every null ray that hits the collapsing object  and reflects off of it will 
reach future
null infinity $\scri^+$. This happens if the null ray collides with the collapsing
object at any stage of the collapse. The only caveat is that the interaction of 
the null ray and the black star will lead to a time delay in the escape of the 
reflected ray but the amount of time delay will depend on when the 
reflection actually occurs.

In contrast to a black star, a ``black hole'' is a classical {\it vacuum} solution 
of Einstein's equations and there is no matter distribution anywhere in
spacetime except perhaps at the central singularity. 
%
Hence the collision of black holes will only 
lead to gravitational radiation because the spacetime is devoid of matter.
On the other hand, black star collisions will lead to gravitational {\it and}
electromagnetic radiation. Collisions of black stars 
can be a source of gamma ray bursts 
({\it e.g.} \cite{Lamb:2000rc,Piran:2004ba}), and such bursts 
will be preceded by gravitational wave emission whose characteristics
are similar to those of black hole mergers. 
Thus, when two black stars collide, they can be distinguished from the
collision of black holes by the presence of electromagnetic radiation. 



In a realistic astrophysical setting the collision of two black holes might be
accompanied by the collision of other accompanying matter, such as an
accretion disk, around the black holes. Stellar mass black holes would 
have devoured surrounding matter and are therefore considered relatively
clean environments, though new astrophysical scenarios might include
such matter \cite{Loeb:2016fzn}. Even if there is surrounding matter that collides 
and produces gamma rays, this electromagnetic radiation and the gravitational
radiation would be produced at the same time, with no specific time delay between
them. On the other hand, the black star scenario clearly predicts a time delay 
between the gravitational wave emission and the gamma ray burst because
first the metric outside the black stars coalesces and only then the material of
the black stars coalesces.


We now estimate the energy radiated from the collision of two black 
stars each of mass $M$. The radius of each of the black stars is 
$R=2GM$ and so the density is 
\begin{equation}
\rho = \frac{1}{8G^3M^2} \sim 10^{14} 
         \left ( \frac{30M_\odot}{M} \right )^2
              {\rm gms / cm^3} 
              \label{bsdensity}
\end{equation}
where $M_\odot \approx 2\times 10^{33}~{\rm gms}$ is the solar mass.
This density is high but still below the QCD density. Hence a $30M_\odot$
black star is mostly composed of ordinary protons and neutrons.

The collision of two black stars of mass $M$ involves the
collision of their constituent protons and neutrons which will lead to 
radiation of photons and other light particles. Since the gravitational
binding between all constituents in this system is very strong, 
we treat the collision as being totally inelastic. Then the initial 
kinetic energy gets converted to radiation resulting in the release 
of total energy
\begin{equation}
E \sim M v^2 \delta \approx 4 \times 10^{49} 
               \left ( \frac{M}{30 M_\circ} \right )
               \left ( \frac{v}{0.5 c} \right ) ^2 \left ( \frac{\delta}{10^{-6}} \right ) ~
                {\rm ergs}
\label{ke}
\end{equation}
where $\delta$ is the gravitational redshift of the energy as it
escapes the collision region.
The collision velocity will typically be an $O(1)$ fraction of 
the speed of light, and the gravitational redshift factor $\delta \ll 1$
will depend on how close the black star is to being a black hole.

We assume that the energy is released in a light crossing
time $\sim R/c = 2GM/c \sim 3\times 10^{-4} {\rm sec}$ (for a
$\sim 30M_\odot$ black star), which is the
only relevant length scale in the problem. This time interval
will be dilated by $\delta^{-1}$ and the emitted power in
photons will be
\begin{equation}
P \sim 2 \times 10^{46} 
         \left ( \frac{v}{0.5 c} \right ) ^2 \left ( \frac{\delta}{10^{-6}} \right )^2 ~
            {\rm ergs/sec}
\label{power}
\end{equation}
Note that the power is independent of the mass $M$. 
We can estimate the frequency of the photons by once again
treating the collision as being totally inelastic. Then every proton in
the black star gets stopped on collision and the emitted photon 
energy is simply the initial kinetic energy of the proton
\begin{equation}
E_\gamma \sim m_p v^2 \delta \approx 0.25
         \left ( \frac{v}{0.5 c} \right ) ^2 \left ( \frac{\delta}{10^{-6}} \right )~ {\rm keV}
\label{Egamma}
\end{equation}


Even though we cannot precisely estimate the event rate of black 
star collisions, we do know that the rate is lower for lower
initial velocity since then the stars will take a long time to collide
{\it i.e.} $\delta$ will be smaller.
Then the emitted energy will redshift by a greater amount and 
there will be a greater time delay between the gravitational and
electromagnetic emissions, making the gamma ray burst very
faint and also temporally uncorrelated with the gravitational wave 
event. On the other hand, we expect that the 
number of black stars falls off with higher velocity. These two 
arguments suggest that there should be a velocity at which black 
star collisions peak. In terms of gamma ray bursts, it implies
that the gamma ray bursts should have a typical photon energy.
Further, the total power emitted should scale with this photon 
energy as seen by dividing Eq.~(\ref{power}) by (\ref{Egamma}),
\begin{equation}
\frac{P}{E_\gamma} \approx 10^{56} ~ 
 \left ( \frac{\delta}{10^{-6}} \right ) ~ {\rm sec^{-1}}
\end{equation}
This formula does not depend on the mass of the colliding
black stars and neither on their velocities, and hence is
an invariant of the model. 




If an observed gamma ray burst is indeed due to colliding 
black stars, the burst should be preceded by gravitational 
wave radiation from the coalescing spacetimes of the black 
stars.  The gravitational wave emission should be very similar 
to that calculated numerically for black hole collisions 
\cite{Pretorius:2005gq,Campanelli:2005dd,Baker:2006yw}, 
and the final gravitational wave emission due to coalescence 
should be accompanied by the gamma ray burst when the material
of the black stars coalesce. 
The waveforms of the emitted electromagnetic radiation will depend 
on the normal modes of the two black star system. Indeed, characteristics 
of the gravitational radiation preceding the gamma ray burst, 
together with the gamma ray burst, may allow us to infer the 
parameters of the colliding black stars and the initial conditions.

LIGO \cite{Abbott:2016blz} has recently detected the gravitational wave signature from
the merger of two black holes, each with mass $\approx 30M_\odot$.
This stunning announcement has been followed by a cautious
but equally stunning claim by the Fermi collaboration \cite{Connaughton:2016umz} 
that they may have seen a gamma ray burst counterpart of the LIGO event.
The energy and emission frequency of the gamma ray burst are broadly
consistent with those estimated for black star collisions. If future gravitational
wave events are followed by delayed gamma ray burst events, it would
be strong support for the black star picture and would provide deep
insight into gravitational collapse, black holes, and quantum gravity.

\begin{acknowledgments}
This work was supported by the U.S. Department of Energy, Office of High Energy Physics,
under Award No. DE-SC0013605 at ASU.
\end{acknowledgments}

\end{document}